# Absence of conventional charge ordering in $Na_{0.5}CoO_2$ from a high resolution neutron diffraction study

A J Williams[1], J P Attfield[1], M L Foo[2] and R J Cava[2]

[1]Centre for Science at Extreme Conditions and School of Chemistry, University of Edinburgh, King's Buildings, Mayfield Road, Edinburgh, EH9 3JZ, United Kingdom

[2]Department of Chemistry, Princeton University, Princeton, NJ 08544, USA

**Abstract**

The structure of $Na_{0.5}CoO_2$, the low temperature insulator that separates the magnetic and superconducting regions in the $Na_xCoO_2 \cdot yH_2O$ phase diagram, is studied by high resolution powder neutron diffraction at temperatures between 10 and 300 K. Profile analysis confirms that it has an orthorhombic symmetry structure, space group *Pnmm*, consisting of layers of edge-sharing $CoO_6$ octahedra in a triangular lattice, with Na ions occupying ordered positions in one-dimensional chains in the interleaving planes. The oxygen content is found to be stoichiometric within 1%, indicating that the Na concentration accurately determines the electronic doping. The Na ordering creates two distinct Co sites with different numbers of Na neighbours, but the difference in their Co-O bond distances and the derived bond valence sums is small.

**Introduction**

Studies of $Na_xCoO_2$ and its hydrated phases have provided a wealth of experimental information on the magnetic and electronic behavior of conducting, planar triangular lattices, and have stimulated considerable theoretical interest. The structures of these materials consist of triangular planes of edge-sharing cobalt-oxygen octahedra, interleaved with either Na or Na-$H_2O$ layers. Characterization of the electronic phase diagram of $Na_xCoO_2$[1] has revealed that an insulating phase, $Na_{0.5}CoO_2$, is found at a doping level between that of the superconductor $Na_{0.35}CoO_2 \cdot 1.4H_2O$[2] and the high thermoelectric coefficient, metallic, magnetic phase $Na_{0.7}CoO_2$[3,4,1]. $Na_{0.5}CoO_2$ displays electronic and magnetic phase transitions at 87, 53, and 20 K[5]. The character of these phase transitions is not fully understood. Although insulating in the comparative sense, the behavior of the resistivity of $Na_{0.5}CoO_2$ is unconventional[1]: the in-plane resistivity increases weakly (from 1 mohm-cm) between 300 and 50K, and then increases sharply, in two steps, at lower temperatures to reach only 20 mohm-cm by 4K. Due to its importance in separating superconducting and magnetic parts of the phase diagram in this system, $Na_{0.5}CoO_2$ is the subject of detailed study (see, e.g. refs 6-11). Based on a conventional neutron diffraction study, and analysis of electron diffraction patterns, it was proposed[5] that the insulating character of this compound was due to the formation of a charge ordered state in the Co lattice induced by the formation of chains of Na running alone one in-plane direction, breaking the hexagonal symmetry and creating two distinct types of Co sites. The quality of the diffraction data, however, was insufficient to characterize the $CoO_2$ lattice in detail. The current study, based on high resolution powder neutron diffraction data, indicates that the structure is not as simply reflective of a conventional charge ordered state as originally proposed, as the structural distinction between the two Co sites is very small. The results support arguments proposed based on NMR and other measurements[6-11] that more subtle effects may be at play. The detailed structural information presented here provides the basis for a deeper understanding of the relationships between the structure and properties of this phase.

**Structure Analysis**

The $Na_{0.5}CoO_2$ sample was synthesised from single phase powder of $Na_{0.75}CoO_2$, itself prepared by heating stoichiometric quantities of $Na_2CO_3$ and $Co_3O_4$ overnight in flowing oxygen at 800°C. That powder was then treated by immersion for 5 days (with stirring) in acetonitrile saturated with $I_2$ in sufficient excess to ensure full oxidation of the powder to $Na_{0.5}CoO_2$. The powder was then washed with acetonitrile and stored in a dry environment during all subsequent handling. Details of the synthesis have been described elsewhere[1].

Powder neutron diffraction data were collected from the High Resolution Powder Diffractometer (HRPD) instrument at the ISIS facility, United Kingdom. Long scans of approximately 6 hours each were collected at 10 and 300 K. Shorter scans of around 1 hour were collected at 30K, 60K and 100K to examine the structure between the three observed transitions in this material. Profiles from the backscattering (2θ=168°) and the 2θ=90° detector banks were simultaneously Rietveld analysed using the General Structure Analysis System (GSAS) program[12]. A linear interpolation background with 10 terms was used, and the peak shape was modelled using a convolution of Ikeda-Carpenter and pseudo-Voigt functions.

A model for the structure of $Na_{0.5}CoO_2$ has been proposed previously [5]. The model has an orthorhombic *Pnmm* symmetry superstructure of the basic hexagonal arrangement, with $a_0 = \sqrt{3}a_H$, $b_0 = 2a_H$ and $c_0 = c_H$, where $a_H$ and $c_H$ are the hexagonal subcell lattice parameters. The departure from hexagonal symmetry was attributed to the ordering of the Na into zig-zag chains running parallel to $b_o$, leading to two different kinds of Co sites. This distinction of Co sites in a 1:1 ratio led to the proposal that a charge ordered state in the Co plane is the origin of the insulating behaviour of $Na_{0.5}CoO_2$. The diffraction data were not of sufficient quality, however, to accurately characterize the cobalt oxygen lattice and confirm the charge ordering.

There are two distinct types of Na site in the $Na_xCoO_2$ family of structures – both are trigonal prismatic, but one shares edges with adjacent $CoO_6$ octahedra (Na2), whilst the other shares faces (Na1). For a broad range of Na contents in $Na_xCoO_2$[13-16], the Na2 site is favoured, however in $Na_{0.5}CoO_2$ it was proposed, and we have found, that the Na1 and Na2 sites are occupied in equal ratios. The highly resolved data from HRPD ($\Delta d/d = 4 \times 10^{-4}$ for backscattering data) allowed for an unconstrained refinement of the lattice parameters and atomic coordinates for $Na_{0.5}CoO_2$ to a stable minimum. The previously proposed model gives an excellent fit to the data at 10 K (Fig. 1) and 300 K, and no improvement was obtained with other superstructure models tested. There is no evidence in the neutron diffraction patterns for any structural distortions to lower symmetry than *Pnmm* down to 10 K. Free refinement of the fractional occupancies at the two Na sites, and at the vacant Na positions, showed that no disorder is present in the Na sublattice to an experimental uncertainty of 2%. Similarly, free refinement of the fractional occupancies at the oxygen sites indicated that no oxygen non-stoichiometry is present, to an uncertainty of less than 1%. This latter result indicates that in the $Na_xCoO_2$ system, the Na content x provides an excellent measure of the formal charge on the Co lattice, and therefore the electronic doping in $Na_xCoO_2$, at least at x=0.5, much as the Sr content allows one to count formal charge and doping in the cuprate superconductor $La_{2-x}Sr_xCuO_4$. The refined structural parameters and selected bond distances are given in Tables II and III.

**Discussion**

The above results confirm that the symmetry of the previously proposed model for the structure of $Na_{0.5}CoO_2$ is correct and that no long-range structural distortions occur on cooling from 300 to 10 K. The weak superlattice observed at 80K by electron diffraction was not observed in the neutron diffraction data. The lattice parameters show a highly anisotropic thermal expansion (Fig. 2), that is consistent with the layered nature of the structure. The averaged in-plane expansion from 10 to 300 K is $\alpha_{ab}$ = (ab/2). $\Delta(ab)/\Delta T$ = 16 ppm, whereas the corresponding value in the perpendicular direction is $\alpha_c$ = 2540 ppm. Comparison of the average Co-O and Na-O distances in Table III, shows that the latter decrease more on cooling from 300 to 10 K, and so the large $\alpha_c$ results mainly from the high thermal expansivity of the height of the $NaO_2$ layers. The volume expansion follows the trend in c, and there is no evidence for a significant volume anomaly between 10 and 300 K (Fig. 2). However, a possible in-plane lattice distortion is evidenced by the temperature variation of the a and b-parameters. A maximum in a is observed at 60 K, although the b- data are more noisy so the expected corresponding minimum is not clear. In the parent hexagonal $NaCoO_2$ structure, the in-plane hexagonal lattice parameter $a_H$ is related to a and b in the present orthorhombic cell by $a_H = a/\sqrt{3} = b/2$. The plot of these normalised cell parameters in Fig. 2 demonstrates that the orthorhombic distortion of the $CoO_2$ planes increases on cooling from 300 to 60 K, but then decreases below 60 K. This suggests that the magnetic ordering and transition to a more insulating state in $Na_{0.5}CoO_2$ at 53K[1,8] may be accompanied by a small structural distortion but without lowering the lattice symmetry.

A partial view of the structure looking along the direction of the Co chains (the b-axis) is shown in Figure 3a. The rows of Na vacancies are clearly seen. Na ordering creates two distinct Co sites, both of which are slightly distorted from regular octahedral geometry (Fig. 3b). The $CoO_6$ geometries do not change significantly between 10 and 300 K. Co(1) shows a small off-centre displacement towards one face of the octahedron (Fig. 3b), while a slight tetragonal elongation (of the Co-O3 bonds) is evident in the $Co(2)O_6$ octahedron. The magnitudes of these distortions are too small for them to signify a local electron-lattice coupling such as orbital ordering at a possible localised $Co^{4+}$ site, in keeping with the Co charge distribution found below

The formal cation charges at the cobalt sites are estimated by the bond valence sum (BVS) method using published parameters[17]. The values in Table III show that Co(1) has a slightly higher formal valence than Co(2), but the difference at 300 K is only 7% of the ideal separation for $Co^{3+}$ and $Co^{4+}$ states, increasing to 12% at 10 K. Recent studies[18-21] have shown that the structural impact of charge ordering is to lead to distinguished bond lengths (and BVS's) of 20-40% of the ideal value in symmetry-broken charge ordered systems, where the structure has been cooled through a transition from a high temperature structure in which all the transition metal sites are equivalent. In $Na_{0.5}CoO_2$, the symmetry equivalence of the two Co sites is broken by the Na ordering, and so charge ordering might have a larger magnitude if the ground state was strongly insulating. In $Mn_2OBO_3$, for example,[22] the ordering of $Mn^{2+}$ and $Mn^{3+}$ over two structurally inequivalent octahedral sites yields essentially 100% of the ideal bond length (or BVS) distinction. The fact that $Na_{0.5}CoO_2$ at high temperatures is only weakly non-metallic and does not undergo a metal to insulator transition that involves a change of orders of magnitude in resistivity is consistent with our observation that the possible separation of charge on the Co(1) and Co(2) sites is not as large as would be expected for a full charge ordered state.

The original proposal for the presence of a charge ordering in $Na_{0.5}CoO_2$ was based on a local charge neutrality picture in which it was postulated that the inequivalent distributions of $Na^+$ ions around the Co(1) and Co(2) sites would tend to result in a lower formal charge for the Co with more Na neighbours in accordance with Pauling's rules. The Na environments around the two Co sites are indeed more different than are their oxygen environments; the Co(1) sites have an asymmetric coordination, with closest Co-Na distances of 2.82 and 3.27 Å, whereas the Co(2) environment is more regular with 2 x 3.19 Å distances. Also, the nearest neighbour $Na^+$ environments - the $Na^+$ ions that are bonded to the oxygens in the $Co(1)O_6$ and $Co(2)O_6$ octahedra (see table IV)- involve more sodium ions for Co(1) than for Co(2), suggesting that the former should be less highly charged to maintain local charge neutrality.

The present structural results suggest, however, that the Na ion ordering leads only to a slight difference in electron density at the two Co sites in $Na_{0.5}CoO_2$, and that spontaneous (or Na order-assisted) $Co^{3+}/Co^{4+}$ full charge ordering is not seen down to 10 K. The further observations that $Na_{0.5}CoO_2$ is surrounded by metallic phases at slightly higher and lower electron counts in the $Na_xCoO_2$ phase diagram, and is only weakly insulating itself, suggests that the Co $t_{2g}$ holes are mainly delocalised even in the x=0.5 phase. It may be that the Na ordering perturbs the Co-O band structure enough to open a small gap at the Fermi level, but without the full charge localisation and ordering that occurs e.g. in manganese oxide perovskites. Alternatively, the broken symmetry induced by the Na ordering may allow more subtle features in the electronic structure to determine the properties, as is presently under discussion in the literature. It is of substantial interest that both magnetic and resistive transitions occur on cooling $Na_{0.5}CoO_2$ and that conventional charge ordering does not appear to be the dominant cause. More detailed study will be of interest to determine whether the small changes in lattice parameters observed in the present study below 50K reflect weak coupling between the subtle electronic and magnetic transitions and the $CoO_2$ lattice in $Na_{0.5}CoO_2$ at low temperatures. Further, a theoretical treatment that addresses the origin of the symmetry breaking Na ordering array in $Na_{0.5}CoO_2$, taking into account the relative energies of the two types of Na positions, the off-center position of the Na in the site that shares edges with the $CoO_6$ octahedra, the Na-Na interactions, possible Na-Co interactions, and the effect of ordering on the electronic density of states, would be of great interest.

**Acknowledgements**

We thank EPSRC for provision of neutron beam time at ISIS, and Dr. R. Ibberson for help with data collection. The work at Princeton was supported by the US department of Energy.

**References**

[1]Foo M L, Wang Y, Watauchi S, Zandbergen H W, He T, Cava R J and Ong N P, *Phys. Rev. Lett*. **92**, 247001 (2003)

[2]Takada K, Sakurai H, Takayama-Muromachi E, Izumi F, Dilanian R A, and Sasaki T, *Nature* **422** 53 (2003)

[3]Terasaki, I., Sasago, Y. and Uchinokura, K., *Phys. Rev*. **B56** R12685 (1997)

[4]Bayrakci S P, Bernhard C, Chen D P, Keimer B, Kremer R K, Lemmens P, Lin C T, Niedermayer C, and Strempfer J, Phys. Rev. B69 100410(R) 2004

[5]Huang Q, Foo M L, Lynn J W, Zandbergen H W, Lawes G, Wang Y, Toby B H, Ramirez A P, Ong N P and Cava R J, *J. Phys. Cond. Mat*. **16**, 5803 (2004)

[6]Hwang J, Yang J, Timusk T, and Chou F C, *Phys. Rev*. **B72** 024549 (2005)

[7]Mukhamedshin I R, Alloul H, Collin G, and Blanchard N, *Phys Rev. Lett.* **94,** 247602 (2005)

[8]Mendels P, Bono D, Bobroff J, Collin G, Colson D, Blanchard N, Alloul H, Mukhamedshin I, Bert F, Amato A, and Hillier A D, *Phys. Rev. Lett.* **94**, 136403 (2005)

[9]Yokoi M, Moyoshi T, Kobayashi Y, Soda M, Yasui Y, Sato M, and Kakumai K, Cond-mat/0506220

[10] Gavilano JL, Rao D, Pedrini B, Hinderer J, Ott HR, Kazakov SM, and Karpinski J, *Phys. Rev*. B**69** 100404 (2004)

[11] Bobroff J, Lang G, Alloul H, Blanchard N, and Collin G, Cond-mat/0507514 (2005)

[12]Larson AC, Von Dreele RB, General Structure Analysis System (GSAS) Report No. LAUR 86-748, Los Alamos National Laboratory (1994) (unpublished).

[13]Delmas C, Braconnier J J, Fouassier C and Hagnemuller P, *Solid State Ion*. **3/4**, 165 (1981)

[14]Fouassier C, Matejka G, Reau J-M and Hagnemuller P, *J. Solid State Chem*. **6**, 532 (1973)

[15]Lynn J W, Huang Q, Brown C M, Miller V L, Foo M L, Schaak R E, Jones C Y, Mackey E A and Cava R J *Phys. Rev. B* **68**, 214516 (2003)

[16]Jorgensen J D, Avdeev M, Hinks D G, Burley J C and Short S, *Phys. Rev. B* **68**, 214517 (2003)

[17]Brese N E and O'Keeffe M, *Acta Crystallogr. B* **47**, 192 (1991)

[18] Williams A J and Attfield J P, *Phys. Rev. B* **66**, 220405 (2002)

[19] Williams A J and Attfield J P, *Phys. Rev. B* **72**, 024436 (2005)

[20] Wright J P, Attfield J P and Radaelli P G, *Phys. Rev. Lett*. **87**, 266401 (2001)

[21] Goff R J and Attfield J P, *Phys. Rev. B* **70**, 140404 (2004)

[22] Norrestam R, Kritikos M and Sjödin A, *J. Solid State Chem.* **114**, 311 (1995)

**Figure Captions**

**Fig. 1** Powder neutron diffraction profiles for $Na_{0.5}CoO_2$ at 10K, fitted in orthorhombic space group *Pnmm*. (a) Backscattering ($2\theta = 168°$) and (b) $2\theta = 90°$

**Fig. 2**(a) Temperature dependence of (upper panel) cell parameters, shown as *a*/√3 (triangles), *b/2* (circles) and *c* (crosses), and (lower panel) cell volume of $Na_{0.5}CoO_2$. Lines are guides to the eye. Error bars are smaller than the points.

**Fig. 3** Views of the $Na_{0.5}CoO_2$ structure (a) $CoO_2$ plane and adjacent Na layers in $Na_{0.5}CoO_2$ viewed along the chain direction, showing the rows of Na vacancies (compared to the parent $NaCoO_2$ type). (b) Co-O distances at 10 K in the linked octahedra, showing the distortions that result from the Na cation ordering. O sites are numbered as in Table II.

**Table I.** Lattice and Agreement Parameters for $Na_{0.5}CoO_2$ at different temperatures.

| T (K):       | 10          | 30          | 60          | 100         | 300         |
|--------------|-------------|-------------|-------------|-------------|-------------|
| a (Å)        | 4.8763(1)   | 4.8764(1)   | 4.8765(1)   | 4.8765(1)   | 4.8761(1)   |
| b (Å)        | 5.6278(1)   | 5.6274(2)   | 5.6276(2)   | 5.6276(1)   | 5.6281(1)   |
| c (Å)        | 11.0620(1)  | 11.0618(2)  | 11.0635(2)  | 11.0694(1)  | 11.1283(1)  |
| V (Å$^3$)    | 303.572(8)  | 303.55(1)   | 303.62(1)   | 303.78(1)   | 305.392(9)  |
| $\chi^2$     | 6.43        | 2.87        | 2.91        | 3.10        | 5.49        |
| $R_{wp}$ (%) | 2.59        | 4.64        | 4.67        | 3.92        | 2.76        |
| $R_p$ (%)    | 3.91        | 7.79        | 7.59        | 6.50        | 4.13        |

**Table II** Refined Atomic Parameters for $Na_{0.5}CoO_2$ at 10 and 300K, in space group Pnmm.

| Atom | Site | T(K) | x | y | z | $U_{iso}$ (Å$^2$) |
|---|---|---|---|---|---|---|
| Co(1) | 4f | 10 | 0.0039(8) | 1/4 | 0.0030(3) | 0.0036(2) |
|  |  | 300 | -0.0021(9) | 1/4 | 0.0028(3) | 0.0067(3) |
| Co(2) | 4d | 10 | 1/2 | 0 | 0 | 0.0036(2) |
|  |  | 300 | 1/2 | 0 | 0 | 0.0067(3) |
| Na(1) | 2b | 10 | -0.0321(6) | 1/4 | 3/4 | 0.0057(3) |
|  |  | 300 | -0.0375(8) | 1/4 | 3/4 | 0.0203(5) |
| Na(2) | 2a | 10 | 0.3609(6) | 3/4 | 3/4 | 0.0057(3) |
|  |  | 300 | 0.3633(8) | 3/4 | 3/4 | 0.0203(5) |
| O(1) | 4f | 10 | 0.3359(4) | 1/4 | 0.0878(1) | 0.0019(4) |
|  |  | 300 | 0.3366(6) | 1/4 | 0.0883(2) | 0.0021(5) |
| O(2) | 4f | 10 | 0.3285(4) | 3/4 | 0.0858(1) | 0.0034(4) |
|  |  | 300 | 0.3298(6) | 3/4 | 0.0850(2) | 0.0074(6) |
| O(3) | 8g | 10 | -0.1616(4) | -0.0013(2) | 0.08890(6) | 0.0062(2) |
|  |  | 300 | -0.1630(5) | -0.0020(3) | 0.08843(8) | 0.0093(3) |

**Table III** Structural results for Na$_{0.5}$CoO$_2$ at 10 and 300K; Co-O distances (Å) and Bond Valence Sums, O-Co-O bond angles (deg.), and Na-O distances.

| Type | | 10K | 300K | Type | | 10K | 300K |
|---|---|---|---|---|---|---|---|
| Co(1)-O(1) | | 1.871(4) | 1.906(5) | Na(1)-O(2) | x2 | 2.321(3) | 2.325(4) |
| Co(1)-O(2) | | 1.895(4) | 1.873(5) | Na(1)-O(3) | x4 | 2.455(2) | 2.477(2) |
| Co(1)-O(3) | x2 | 1.885(3) | 1.880(3) | <Na(1)-O> | | 2.410(2) | 2.426(3) |
| Co(1)-O(3) | x2 | 1.893(3) | 1.904(3) | | | | |
| <Co(1)-O> | | 1.887(2) | 1.891(2) | | | | |
| **BVS** (Co(1)) | | 3.43 | 3.40 | | | | |
| | | | | | | | |
| Co(2)-O(1) | x2 | 1.888(1) | 1.892(1) | Na(2)-O(1) | x2 | 2.325(3) | 2.319(4) |
| Co(2)-O(2) | x2 | 1.892(1) | 1.888(2) | Na(2)-O(3) | x4 | 2.474(2) | 2.489(2) |
| Co(2)-O(3) | x2 | 1.921(2) | 1.915(2) | <Na(2)-O> | | 2.424(2) | 2.432(3) |
| <Co(2)-O> | | 1.900(1) | 1.898(1) | | | | |
| **BVS** (Co(2)) | | 3.31 | 3.33 | | | | |
| | | | | | | | |
| O(1)-Co(1)-O(3) | x2 | 96.8(1) | 96.3(2) | O(1)-Co(2)-O(2) | x2 | 96.3(1) | 96.3(1) |
| O(1)-Co(1)-O(3) | x2 | 85.3(1) | 84.3(2) | O(1)-Co(2)-O(2) | x2 | 83.7(1) | 83.7(1) |
| O(2)-Co(1)-O(3) | x2 | 94.0(1) | 94.7(2) | O(1)-Co(2)-O(3) | x2 | 95.9(1) | 95.7(1) |
| O(2)-Co(1)-O(3) | x2 | 84.0(1) | 84.7(2) | O(1)-Co(2)-O(3) | x2 | 84.1(1) | 84.3(1) |
| O(3)-Co(1)-O(3) | x1 | 97.2(2) | 97.9(2) | O(2)-Co(2)-O(3) | x2 | 96.9(1) | 96.6(1) |
| O(3)-Co(1)-O(3) | x2 | 83.7(1) | 83.9(1) | O(2)-Co(2)-O(3) | x2 | 83.1(1) | 83.4(1) |
| O(3)-Co(1)-O(3) | x1 | 95.3(2) | 94.3(2) | | | | |

**Table IV** Nearest neighbour $Na^+$ ion distances for the Co(1) and Co(2) sites in $Na_{0.5}CoO_2$ at 10 and 300K. Nearest neighbour $Na^+$ are bonded to the oxygens in the $Co(1)O_6$ and $Co(2)O_6$ octahedra.

| Type | Number | 10K | 300K |
| --- | --- | --- | --- |
| Co(1)-Na(1) | x1 | 2.805(3) | 2.819(4) |
| Co(1)-Na(2) | x1 | 3.260(4) | 3.266(5) |
| Co(1)-Na(1) | x2 | 3.924(2) | 3.940(3) |
| Co(1)-Na(2) | x1 | 4.130(4) | 4.155(5) |
| Co(1)-Na(2) | x2 | 4.334(3) | 4.360(3) |
| | | | |
| Co(2)-Na(2) | x2 | 3.176(1) | 3.188(1) |
| Co(2)-Na(1) | x2 | 3.851(2) | 3.848(2) |
| Co(2)-Na(1) | x2 | 4.045(2) | 4.073(3) |

Fig. 1

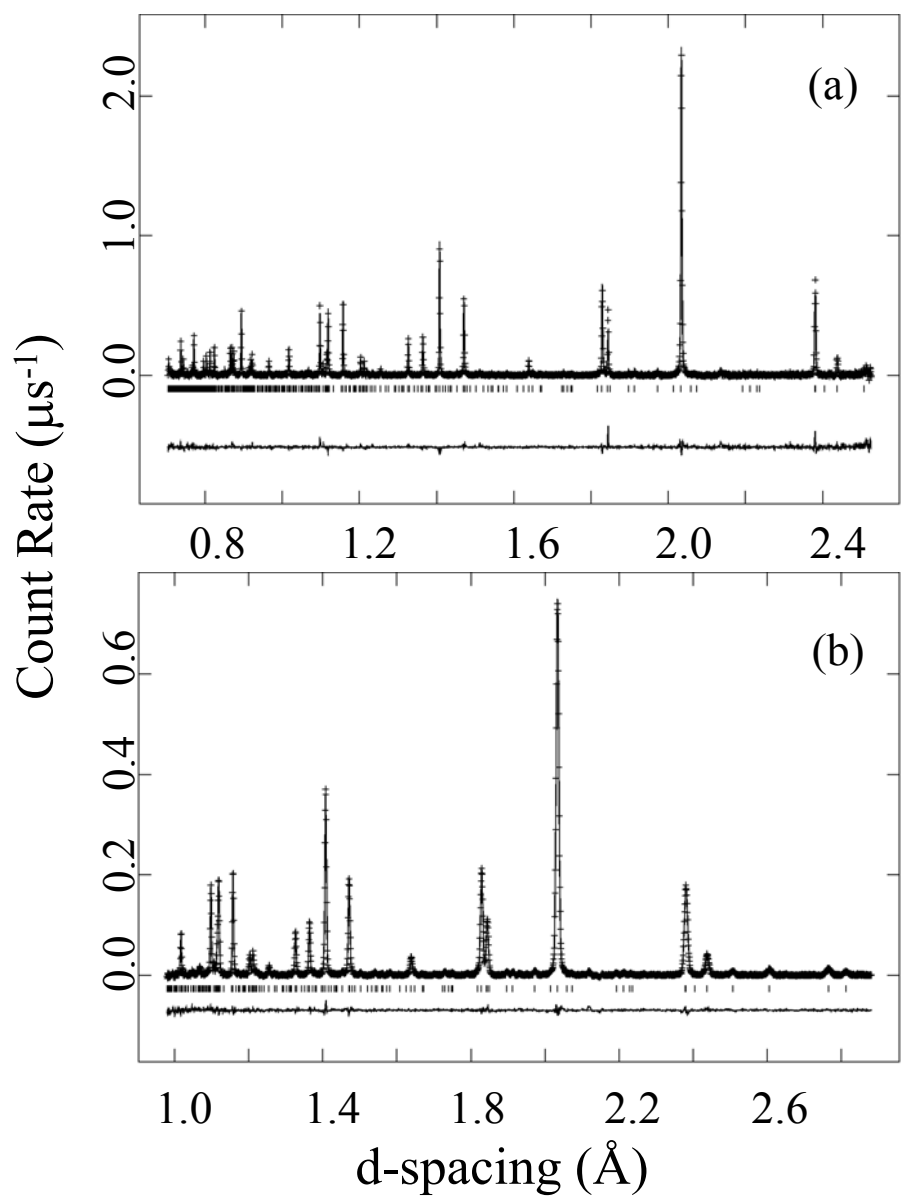

Fig. 2

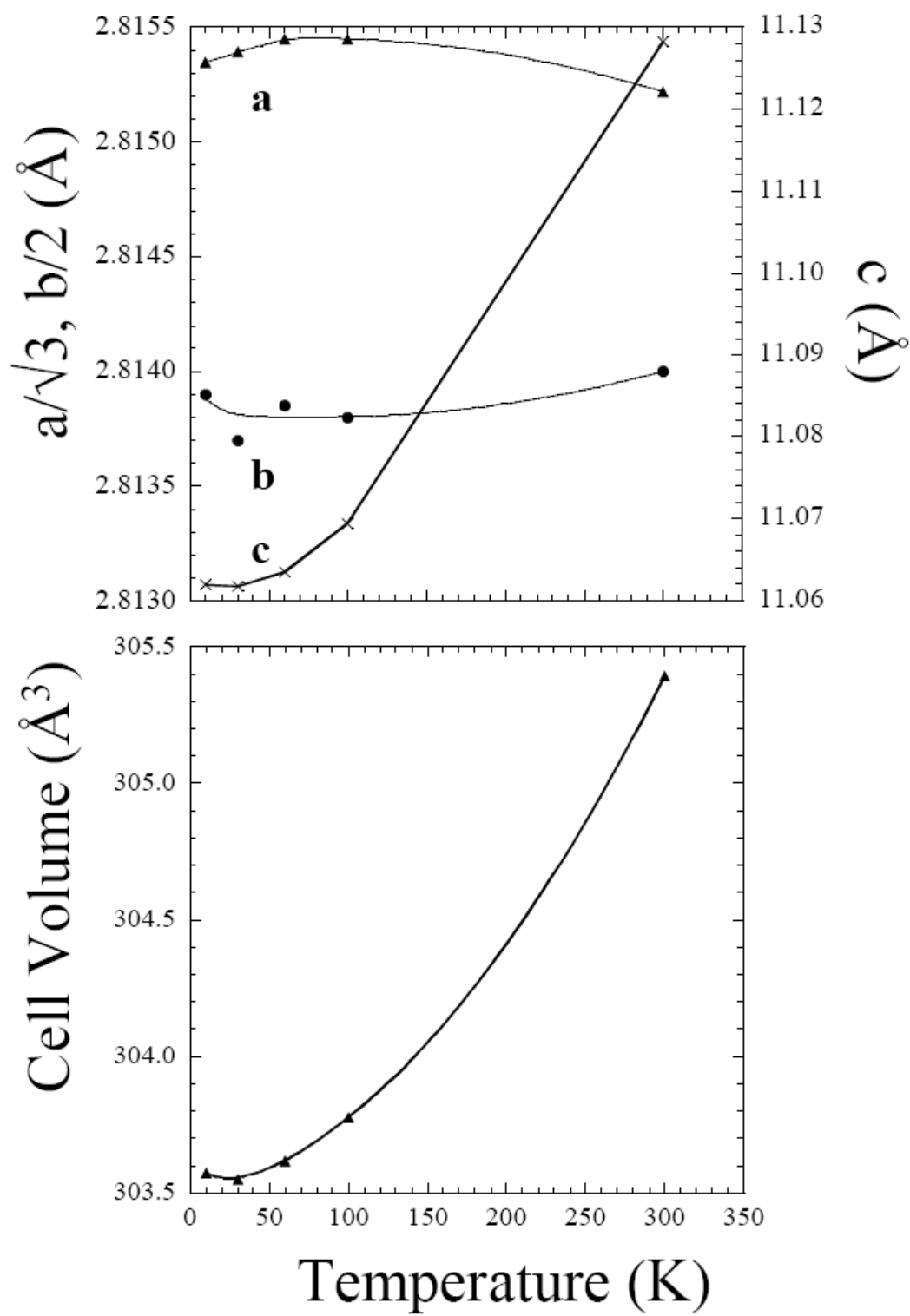

Fig. 3(a)

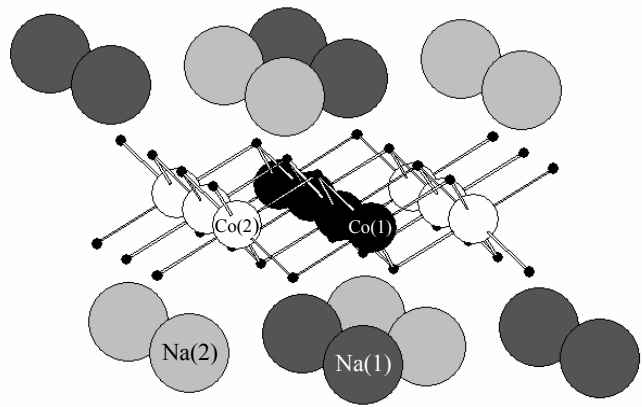

Fig. 3(b)

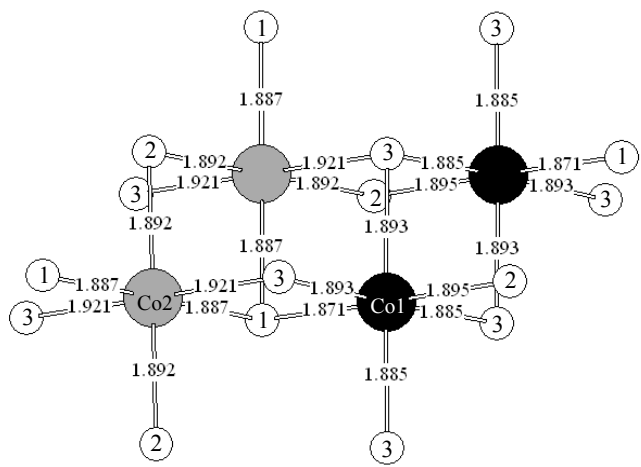